\documentclass[preprint, 12pt]{aastex}
\usepackage{epsfig}

\def\mr{\mathrm}

\begin{document}

    \title{Why the Salpeter screening formula cannot be applied in 
    the Sun}
     \author{Giora Shaviv}
     \email{gioras@physics.technion.ac.il}
     \affil{Department of Physics  and Asher Space Research Institute\\
     Israel Institute of Technology \\
     Haifa, Israel 32,000}
\begin{abstract}
In a recent paper,  Bahcall et  al.  (2000) list various new
approaches to the problem of screening of nuclear reactions in 
stellar plasma and assert that they are all wrong or irrelevant. 
Except for two, all approaches mentioned by Bahcall et al. assume the 
mean field approximation. The two exceptions are Carraro et al. 
(1988) and Shaviv \& Shaviv (2000a). While Carraro et al. (1988) paper is 
discussed shortly and refuted by Bahcall et al. (2000) the Shaviv \& Shaviv 
(2000a) paper is 
not discussed and refuted only by association. However, the association is 
totally unfounded because Shaviv \& Shaviv (2000a) have shown that kinetic 
equations must be used to solve the screening problem and that the 
mean field approximation is inadequate for this problem. They also 
showed that the Carraro et al. (1988) approach is erroneous. 

Therefore we summarize here the method of $S^{2}$ and their main result.
 We contrast  the kinetic equations method with the mean field approximation 
 and  expose the different assumptions and omissions  in each method. 
\end{abstract}

\section{Introduction}
In view of the controversy about how to calculate the screening and 
the new paper by Bahcall et al. (2000) we feel that an explanation of the 
method of Shaviv \& Shaviv (2000a) (hereafter $S^{2}$) and  in 
particular juxtaposition of it with the mean field methods is 
appropriate.
The issue of this paper is not whether the screening resolves or 
not  the solar neutrino problem. This paper is about how to calculate 
the screening irrespective of the consequences to the solar neutrino 
problem. It may very well be that the new screening aggravates the 
classical solar neutrino problem and enhances the discrepancy between 
the prediction of the standard solar neutrino model and the 
experiments.

The plasma correction to the rate of the solar nuclear reactions
 affects the theoretical prediction of the solar neutrino fluxes and
 consequently the predicted counting rate in the various undergoing
 experiments to detect solar neutrinos.  Therefore, as accurate values
 as possible are needed for evaluation of the nuclear reaction
 rates.  As is well known, if all attempts to explain the solar
 neutrino discrepancy between theory and observations fail, then one
 of the suggested explanations is neutrino oscillations.  In this
 case, the accurate derivation of the parameters of the solution, like
 the oscillation length, would depend on the exact prediction of the
 neutrino fluxes in the classical theory.
{\it As plain as day, before a new theory is invented it is crucial to 
calculate correctly all effects created  by the `classical' theory. }

The paper by Bahcall et al. (2000) enumerates various attempts to derive the 
screening correction to the rate of nuclear reactions in stellar plasma. 
The paper gives reference to  $S^{2}$ in the context of 
dynamic screening and then does not discuss the paper, the method or 
the results,  but instead 
discusses the dynamic screening of Carraro et  al. (1988). As a mater 
of fact, $S^{2}$ show  
in detail that Carraro et al. (1988) are wrong in using the approach  of a 
test particle instead of a particle in thermodynamic equilibrium. 
Neither the effect $S^{2}$ discuss, nor their kinetic approach,  are mentioned 
in the Bahcall et al. (2000)
paper but the impression the reader is provoked to adopt is that it is 
identical to Carraro et al. and hence equally wrong (see section 3.1 of their paper).
This is far from being the truth and the purpose of 
this paper is  therefore
\begin{itemize}
    \item  To show  in  what way the method of $S^{2}$ is new and 
different from the mean field methods.
    \item  To explain what is the plasma effect $S^{2}$ discuss
and why it is a natural consequence of an equilibrium state.
     \item  Explain why the assumption of mean field does  not 
apply to the screening in the Sun. 
\end{itemize}
We first discuss the premises of the standard treatment of screening 
and then enumerate the assumptions of the $S^{2}$ method, as well as 
the $S^{2}$ definition of  the 
screening from first principles and finally compare the various sets 
of tacit assumptions in each approach.

\section{The standard  treatment of screening}
All  treatments of the screening (except for Carraro et al. 1988 and 
$S^{2}$) are 
based on the mean field approximation.  The mean field  is 
the {\it average} field a particle feels in the plasma. The average 
is calculated over a thermodynamically long times. According to the 
ergodic hypothesis this field is equal to the average field calculated 
in a snapshot with the average taken over all particles in the system.
Is this approximation valid for 
the solar conditions? The Debye radius in the solar core is about 
$0.87<r>$, where $<r>$ is the mean interparticle distance,
so that  $N_{D}$, the mean number of particles in a Debye 
sphere, is 
2-5 (depending on the charge of the ion). The mean field 
approximations 
treats this number as a fixed and very large number which does not fluctuate
(because it is assumes that $1/\sqrt N \ll 1$) so as 
to obtain the mean field potential.  The first claim of $S^{2}$ is that 
because of the smallness of $N_{D}$ the mean field approximation is 
a poor approximation for reaction kinetics. (cf Montgomery \& Tidman 
1964 for a discussion of the breakdown of the cluster expansion when 
$N_{D}$ is small)

Why do the fluctuations matter? When the relevant property is constant
with the kinetic energy  like the potential energy, the averaging 
over the fluctuations is equivalent to averaging over a constant mean 
potential. But when a phenomenon is sensitive to the 
energy  like the Coulomb barrier penetration,   the average over  fluctuations 
is not equivalent to penetration with the average energy.

The Bahcall et al. (2000) refers to the screening in the kinetic 
approach a la Clayton (1968) where particles react via a mean 
potential. However, we mention in passing that
one can discuss the screening 
effect through a change in the number density of particles due to the 
correlation (cf Ichimaru (1994)) without considering the interaction 
directly. 

\section{The  treatment of $S^{2}$}
$S^{2}$ assume neither  the mean field nor that the amount 
of energy transferred in a ion-ion scattering is the mean potential energy 
per ion, but instead 
start from first principles. They defined the plasma effect 
during a scattering of two ions and calculated it {\it directly} without 
any additional assumption about the long time average field.

 The formalism to include the plasma effect on the
 nuclear rate   assumes that the particles are free at infinity and
 the only interaction is through the bare pair potential. In other 
 words, the
 classical stellar nuclear reaction theory defines the screening
 energy as follows. The total energy of a pair (ignoring the
 surrounding plasma) is given as:
\begin{equation}
E_{\mr{bin}}^{\mr{pair}}=E_{\mr{kin,1}}+E_{\mr{kin,2}}+{{Z_1Z_2e^2} \over {r_{12}}}.
    \label{eq:pair-ener}
\end{equation}
The screening energy, that is the energy  the incoming pair gains 
from the plasma, is then given by:
\begin{equation}
E_{\mr{scr}} =\Delta E_{\mr{bin}}^{\mr{pair}}=E_{\mr{bin,c}}^{\mr{pair}}-E_{\mr{bin,f}}^\mr{pair}
    \label{eq:screen-def2}
\end{equation}
 where index c means close and index f means far away.  Equation
 \ref{eq:screen-def2} compares directly the energy of the pair when it
 is close and when it is far away.  The calculation should proceeds 
 therefore,
 as follows:   calculate the evolution of all particles in the
 system.  For every particle, find the nearest particle and  declare 
 it as `mates'.  Evaluate now the dynamic evolution from the
 identification moment as a pair of approaching particles, through the
 approach and until the pair separates a given distance.  Once the 
 particles moved away calculate Eq. \ref{eq:screen-def2}. In 
 this way the screening is calculated directly from first principles 
 without any additional assumptions. 

 Clearly, when the reaction takes place in vacuum $E_\mr{scr} \equiv
 0$.  However, any effect the plasma has on the approaching pair will
 appear in $E_\mr{scr}$. We stress that Eq.~\ref{eq:screen-def2}
 correlates between the distance of closest approach and far away and
 not between the `initial' and `final' scattering states. According to
 the classical Salpeter  or any mean field theory, there is no difference between the
 above two:  the energy gained from the plasma by the
 approaching particles is returned by the separating particles. The
 balance is maintained per each collision.

 In summary, the mean potential energy per particle, $E_\mr{pot}$, namely
 the long time average of the potential energy, does not depend on the
 absolute or the relative kinetic energy of the particle. This
 classical result of statistical mechanics is manifested clearly in
 the calculations of $S^2$.  However, the energy gained (or lost)
 from the plasma by a scattering pair as given by
 Eq. \ref{eq:screen-def2} may depend on the relative kinetic energy.

 It is clear that in order to solve Eq. 2 for particles in the plasma 
 one needs a proper kinetic treatment and not a mean field approach. $S^{2}$ 
 found 
 that in view of the complexity of the problem (see also later) it is 
 advantageous to look for a method that handles the problem from 
 first principles and without any additional assumptions or 
 approximations. Such a method is the Molecular Dynamics method (MD).

 What is the Molecular Dynamic method? One takes a system of N 
 positive (the protons) and N negative (the electrons) particles, 
 assumes a pure Coulomb  interaction between every two 
 particles (p-p, p-e and e-e) and solves the 3N second order Newton's 
 equation of motion. The MD method does not assume anything about a 
 mean field or smoothing field etc. Nor  does it assume a long time 
 average potential for the scattering of any two particles. In the MD method 
 the scattering of each pair with all the interaction of the 
 particles around it is exactly  followed. To have a decent representation 
 of the various effects N must be large. In the case of $S^{2}$ 
 $N=10^{5}$ was used. (Some authors apply  screened potential. However, $S^{2}$ 
 apply a simple $1/r$ potential).

\subsection{The new effect}
 Consider a pair of mutually scattering particles with a given
 relative kinetic energy between them. The basic findings of S$^2$ 
 is that when the relative kinetic energy of the pair is low it
 gains energy from the plasma as they approach each other. On the 
 other hand, particles with large relative
 kinetic energy lose energy to the plasma as they approach each other.
 Gain or loss are used in
 terms of the relative kinetic energy of the pair and not in absolute
 terms.  If one takes a random pair of approaching particles with a 
 given relative kinetic energy, it may 
 lose or gain energy. However, the average for a given relative 
 kinetic energy is positive for low relative kinetic energies and 
 negative for high relative kinetic energies. ({\it In this respect it  is exactly like 
 the action of 
 dynamic friction in a cluster of stars. }). 
 
 Of course, on the average the energy gained/lost by the
 scattering particles when summed overall particles in the plasma,
 {\bf must } vanish. The balance is over all particles not per 
 collision. 
 
 The physical explanation of the effect is simple.
 When the relative energy of the
 scattering particle is lower than the mean thermal energy of the
 particles in the screening cloud, it gains energy upon penetrating
 into the cloud.  However, when the relative kinetic energy is higher
 than the mean thermal energy, the pair loses energy by penetrating
 into the cloud of each other.  One could predict the existence of 
 such an effect without going into the long calculations. However, the 
 amount of energy exchange and the energy of turn over from gain to 
 loss must be obtained from calculations. 
 
 We note at this point that none of 
 the various treatments of screening (for reference see Bahcall et 
 al.  2000) satisfy the condition of overall energy balance explicitly. In other words,
 how a pair of 
 scattering protons, which gains  
 energy from a cloud of 3 particles upon approaching each other, 
 returns this energy to the cloud 
 as they separate?
 The classical treatments assume implicitly  a detailed balance, namely each 
 approaching pair, which gains energy from the plasma, must return it to 
 the plasma upon separation (which is apparently good for 
 $N_{D}\rightarrow \infty$). If this assumption is dropped from the 
 mean field approximation, then one has to explain how the 
 approaching particles return the gained energy to the plasma.

 \subsection{The MD method reproduce the statistical mechanics 
 results }
 Without fail the MD reproduces all standard statistical mechanics 
 results obtained after a long time average. The effect found in 
 $S^{2}$ does not violate thermodynamics nor does it need a new or 
 special assumption about thermodynamics or screening. It is a simple 
 consequence of the global balance in the plasma. If some particles 
 gain energy from the plasma others must lose energy to the plasma.

The  MD calculation of $S^{2}$ reproduces the 
following standard statistical mechanics results:
\begin{itemize}
    \item  {The particles obey a Maxwell-Boltzmann distribution. The 
    calculation starts from an arbitrary initial distribution in the 
    phases space
    and relaxes after an initial time to a very accurate Maxwell-Boltzmann
    distribution. The distribution is reproduced over many 
    orders of magnitudes in number of particles.}

    \item  {The long time average of the potential energy per particle 
    does not depend on the absolute kinetic energy of the particle.} 

    \item  {The average force acting on the particle vanishes but the 
    root mean square does not and as a matter of fact is very large and does not 
    depend on the kinetic energy of the particle.}

    \item  {The potential energy of the particle does not depend on 
    the mass of the particle. The potential energy per particle does 
    not change when the mass of the particles changes.}

    \item  {The long time average potential between two ions, say 
    protons, is the Debye H${\ddot {\rm u}}$ckel potential and the 
    potential energy per particle is close to $\Gamma$, the plasma 
    parameter. }
    
    \item{All thermodynamic properties like mean kinetic 
    energy, mean potential energy per particle etc are in agreement 
    with statistical mechanics.}

    \item{The power spectrum of the fluctuations does not show any scale and 
    obey a power law.}
    
    \item{The distribution of the potential energy per particle as 
    seen in a snapshot (space average) is properly obtained. }
\end{itemize}

    \subsection{What MD has that mean field does not}
What does the Molecular Dynamic  contain that the mean field is 
missing:
\begin{itemize}
    \item  {The exact energy exchange  between  the two approaching 
    ions and {\it each} massive ion in the 
    cloud  around the interacting pair is followed. 
The same is true of the energy exchange of the approaching pair and 
the electrons.     Thus, while the pair of approaching particles 
gains/loses energy, the energy lost/gained by the other particles is 
fully accounted for. 
    In the mean field approach the energy lost by the 
    cloud of particles composing the mean field is not accounted 
    for.  The implicit assumption is that the mass of the potential 
    is infinite. }

    \item {The recoil of each particle composing the `effective 
    potential' upon the approach of the interacting pair of ions is 
    fully accounted for in the MD method.  In 
    the mean field approach the tacit assumption is that the potential 
    has an infinite mass and does not recoil (or lose/gain energy 
    during the approach). }

    \item  {The MD secure automatically an overall detailed balance
    of the energy exchange between the 
    plasma and the scattering pair. Some pairs lose energy while 
    others gain, but the sum over the entire system vanishes. 
     In the mean 
    field one assumes a detailed balance, namely while a pair of ions 
    gains energy from the plasma as they approach each other, it
    loses {\it exactly } this  energy as it moves away. But it is 
    difficult to see how 
    energy gained from a cloud of 2-5 particles during the approach 
    is exactly returned  to the cloud during the separation, or is it returned 
    elsewhere? }

    \item  {As the number of particles in the Debye sphere is very 
    small, the MD treats properly the third body 
    interaction  as well as all higher orders. In the mean 
    field, it is always the effect of the 'many' particles composing 
    the mean spherical field. }

    \item  {On the one hand the instantaneous potential energy of all particles 
    is not identical  (it fluctuates) and there is a distribution of 
    potential energies.  On the other hand, the Coulomb penetration 
    is very not linear with kinetic energy. Consequently, taking the Coulomb 
    penetration at the value of the mean field may be very inaccurate. }
    
    \item{ The mean field approach assumes that the energy gained by 
    the approaching pair (and later returned to the plasma) 
     is always the mean potential energy of an ion in the plasma. The 
     MD calculates the energy exchange exactly without recourse to any 
     assumption.}
     
     \item{The method of Molecular Dynamics is equivalent to summing 
    the  ladder (interaction) diagrams to all  orders.}

\end{itemize}

\section{Conclusions}
\subsection{General conclusions}
\begin{itemize}
    \item{$S^{2}$ applied Molecular Dynamics to calculate from {\it first 
    principles} the effect of 
    the plasma on a pair of reacting particles. The numerical method 
    is equivalent to summation of the interaction diagrams  to all orders.}
    \item{The MD reproduces all results of statistical mechanics to a 
    high accuracy.}
   \item  The basic result of $S^{2}$ 
   is that the mean energy exchange with the plasma is positive for 
    pairs with low relative kinetic energy (the pair gains energy)  and 
    negative for high relative energy 
    pairs (the pair loses energy).  The sum overall scatterings
      vanishes for each specie in the plasma.
   \item  The interaction of the plasma with the scattering particles 
    is dynamic and must be derived from a proper kinematic considerations.
   \item  The spread in the energy exchange of the particles with 
    the plasma is large and must be taken into account in calculating 
    the effective Coulomb penetration. 
      \item The pair-plasma energy exchange in a single scattering 
      is
    mainly  between the massive pair and the  massive ion component.  The 
    contribution of the electrons in a {\it single} collision is 
    calculated by the MD method but is effectively 
    negligible.  Of course, over thermodynamic times there is an 
    energy exchange between the massive ions and the light electrons. 
    \item The mean field approximation ceases to be valid under the 
    conditions prevailing in stellar cores in general and in the Sun 
    in particular. Even using only the ion contribution to the mean 
    field is not sufficient since few body interactions cannot 
    be neglected.  
    
    \item{Averaging the effect of the plasma into a mean 
    long term potential is not appropriate for handling very non 
linear functions, like the Coulomb barrier penetration.}

    \item  The penetration factor calculation must include the spread in the 
    energy exchange. 
    \end{itemize}

    \subsection{Could the screening affect the prediction of the solar 
    neutrino  flux?}
   While it is not the purpose of this paper to discuss this 
   particular problem we would like to comment, as an epilogue, as follows: all 
   numerical tests about the effect of the plasma assumed a mean 
   field approximation where all reaction are either enhanced or not 
   affected (for a review see Dzitko \& Turck-Chi\`eze(1995).  
   Shaviv \& Shaviv (2000b) show that (a) due to 
   fluctuations the problem is 
   more complicated than just assumed by the approach using  a `corrected' penetration 
   factor. (b) Some reactions are enhanced and others are suppressed 
   depending on the relation of the Gamow energy to the turnover energy 
   (from gains to losses) and the width of the screening distribution.  
   Finally, assessment of the effect is carried out and will be 
   published in due course. The effect found so far is far from being 
   negligible. 
   
   \section{Acknowledgment}
   The author  would like to acknowledge extremely helpful discussions with J.P. Lasota.
   Also, he  thanks the Institut d'Astrophysique de Paris for hospitality and the
   European Association for Research in Astronomy for supporting
his visit there.

\end{document}